\documentclass[aps,prl,letterpaper,nofootinbib,floatfix,12pt]{revtex4}

\usepackage{amsfonts}
\usepackage{amsmath}
\usepackage{amssymb}
\usepackage{graphicx}
\usepackage{comment}
\usepackage{times}
\usepackage{setspace}

\begin{document}

\singlespacing

\title{Influence of the environment and probes on rapid DNA sequencing via
transverse electronic transport}

\renewcommand{\thefootnote}{\fnsymbol{footnote}}
\author{$^{1}$Johan Lagerqvist, $^{2}$Michael Zwolak, and $^{1}$Massimiliano Di
Ventra\footnote{E-mail address: diventra@physics.ucsd.edu}}
\affiliation{$^{1}$Department of Physics, University of California,
San Diego, La Jolla, CA 92093-0319} \affiliation{$^{2}$Physics
Department, California Institute of Technology, Pasadena, CA 91125}
\date{\today}
\begin{abstract}
We study theoretically the feasibility of using transverse
electronic transport within a nanopore for rapid DNA sequencing.
Specifically, we examine the effects of the environment and
detection probes on the distinguishability of the DNA bases. We find
that the intrinsic measurement bandwidth of the electrodes helps the
detection of single bases by averaging over the current
distributions of each base. We also find that although the overall
magnitude of the current may change dramatically with different
detection conditions, the intrinsic distinguishability of the bases
is not significantly affected by pore size and transverse field
strength. The latter is the result of very effective stabilization
of the DNA by the transverse field induced by the probes, so long as
that field is much larger than the field that drives DNA through the
pore. In addition, the ions and water together effectively screen
the charge on the nucleotides, so that the electron states
participating in the transport properties of the latter ones
resemble those of the uncharged species. Finally, water in the
environment has negligible direct influence on the transverse
electrical current.
\end{abstract}

\maketitle

\markright{DNA sequencing via transverse transport}
\section{INTRODUCTION}
Now that the first full human genome has been
sequenced~\cite{Lander2001-1,Venter2001-1}, new uses of sequencing
in medicine seem to be on the horizon. One of the most ambitious
goals is to be able to sequence an entire human genome in less than
an hour for about 1,000 USD, allowing for every-day sequencing in
medicine.~\cite{Collins2003-1}

Several intriguing sequencing
methods~\cite{Kasianowicz1996-1,Braslavsky2003-1,Zwolak2005-1,Gracheva2006-1,Lagerqvist2006-1}
have been proposed which would lead us closer to achieving this
goal. Many of these methods are based on the idea of translocating
DNA through a nanopore.~\cite{Kasianowicz1996-1,
Akeson1999-1,Deamer2000-1,
Meller2000-1,Vercoutere2001-1,Meller2001-1,Deamer2002-1,Meller2002-1,Li2003-1,
Nakane2003-1,Aksimentiev2004-1,Chen2004-1,Fologea2005-1,Heng2006-1,Li2001-1,
Storm2003-1,Harrell2003-1,Li2004-1,Lemay2005-1} In their pioneering
work Kasianowicz {\it et al}. demonstrated that DNA can be pulled
through a biological nanopore roughly the size of the DNA
itself.~\cite{Kasianowicz1996-1} The translocation of the DNA can be
detected by measuring a blockade current when ions are partially
prevented from entering the pore. More recent experiments have been
based on solid state pores made of silicon-based
materials.~\cite{Li2001-1,Storm2003-1,Harrell2003-1,Li2004-1,Lemay2005-1,Mannion2006-1,
Biance2006-1,Ji2006-1} The advantage of solid state pores is that it
may be possible to embed single molecule sensors in the pore to
measure various physical properties of the DNA during translocation,
allowing the DNA to be directly sequenced by detecting specific
signatures of individual bases.

\begin{figure}
%\begin{center}
\includegraphics*[width=7.5cm]{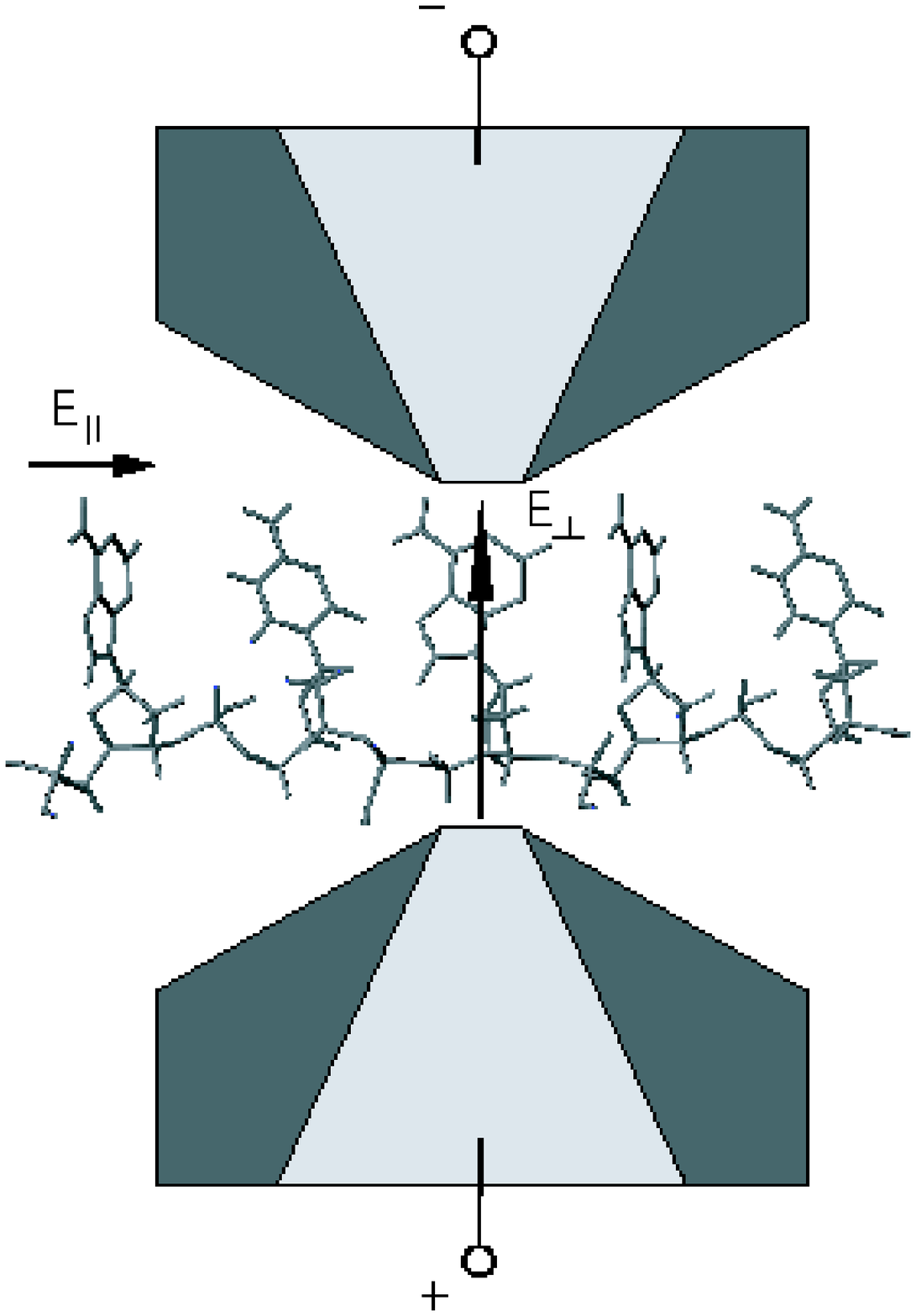}
\caption{} \label{schematic}
%\end{center}
\end{figure}

Previous work has shown the potential for sequencing DNA by
measuring a transverse electronic current~\cite{Zwolak2005-1} as
single-stranded DNA (ss-DNA) translocates through a
nanopore.~\cite{Lagerqvist2006-1} The concept envisions a nanopore
device with embedded nanoscale gold electrodes\footnote{In these
proof-of-concept calculations, we assume gold electrodes. One can
also envision electrodes made out of other materials, such as carbon
nanotubes. Change of material will, for example, change the coupling
in between the DNA and electrodes. The main conclusions drawn in
this and previous work~\cite{Lagerqvist2006-1} will however not
change as the calibration of the device (as discussed later) will
take into account the microscopic details of the nanopore and
electrodes.}, as schematically shown in Figure~\ref{schematic}.
Operating such a device with a transverse field, $\mathbf{E_\perp}$,
(the field that drives the electronic current) greater than the
longitudinal pulling field, $\mathbf{E_\parallel}$, (i.e., the field
that drives DNA translocation) stabilizes the motion of the
nucleotide between the electrodes.~\cite{Lagerqvist2006-1} This
creates a very desirable situation where structural fluctuations
(the most important source of intrinsic
noise~\cite{Lagerqvist2006-1}) are reduced to such a level that {\em
distributions} of currents for each base, while still overlapping,
are different enough to allow for high statistical
distinguishability between the different bases. In previous work,
however, it was assumed that each measurement could be performed
almost instantaneously. This is just a theoretical assumption, and a
more realistic treatment of the measurement probes needs to be taken
into account.

In this paper we will examine the fact that, contrary to naive
expectation, the measurement bandwidth of the electrical probes
reduces these overlapping distributions into sharply peaked and
disjoint distributions, rather than just limiting the sampling rate.
Therefore, assuming that no external sources of noise are present
other than shot, thermal and structural fluctuation noise, a single
current measurement may be sufficient to distinguish each individual
base. Thus by measuring the current as the nucleotides translocate
through the pore, the DNA may be accurately sequenced in extremely
short time scales.

In ref.~\onlinecite{Lagerqvist2006-1} it was estimated that the raw
sequencing throughput of a single 12.5 $\textrm{\AA}$ pore,
operating with a 1V transverse bias, could be as high as 3 billion
bases in 7 hours. In a real device, however, the pore diameter will
not be easy to control and it may not be possible to operate the
device at high transverse biases. Also, ions and water in the pore
are additional sources of noise on top of the structural
fluctuations of the ss-DNA. In this article we will examine in
detail these effects on the distinguishability of the bases. To do
this, we use all-atom molecular dynamics simulations coupled with
quantum mechanical current calculations. We find that although the
overall magnitude of current can change dramatically, the intrinsic
distinguishability of the bases is not significantly affected by
pore size and transverse field strength. The latter is the result of
very effective stabilization of the DNA by the transverse field,
$\mathbf{E_\perp}$, so long as that field is much larger than the
pulling field, $\mathbf{E_\parallel}$. In addition, the ions and
water together effectively screen the charge on the nucleotides, so
that the electron states participating in the transport properties
of nucleotides in solution resemble those of uncharged species.
Finally, water in the environment has a negligible direct influence
on the electrical current through the DNA.

\section{SETUP AND METHODS}

To calculate the current, we use a scattering approach with a
tight-binding Hamiltonian to represent the electronic structure of
the system. For each carbon, nitrogen, oxygen, and phosphorus atom
$s-, p_x-, p_y-$, and $p_z-$ orbitals are used, while $s$-orbitals
are used for hydrogen and gold. We take the Fermi level to be that
of bulk gold, which is identical to that of the extended molecule.
However, for the biases we consider, the current calculations are
relatively insensitive to the exact position of the Fermi level, as
it falls within the highest occupied molecular orbital (HOMO)-lowest
unoccupied molecular orbital (LUMO) gap.~\cite{Lagerqvist2006-1} The
retarded Green's function, $\mathcal{G}_{DNA}$, can be written as
\begin{equation}
\mathcal{G}_{DNA}(E)=[E\mathcal{S}_{DNA}-\mathcal{H}_{DNA}-\Sigma_t-\Sigma_b]^{-1},
\end{equation}
where $\mathcal{S}_{DNA}$ and $\mathcal{H}_{DNA}$ are the overlap
and the Hamiltonian matrices and $\Sigma_{t(b)}$ are the self energy
terms that describe the interaction in between the leads and the
DNA. For a given $\mathcal{G}_{DNA}$ the transmission coefficient
can then be calculated as
\begin{equation}
T(E)=\mathrm{Tr}[\Gamma_t\mathcal{G}_{DNA}\Gamma_b\mathcal{G}_{DNA}^\dagger],
\end{equation}
where $\Gamma_{t(b)}=i(\Sigma_{t(b)}-\Sigma_{t(b)}^\dagger)$.
Finally the current in between two electrodes is given by
\begin{equation}
I = \frac{2 e}{h} \int_{-\infty}^{\infty} dE T(E) [f_t(E) -f_b(E)]
\end{equation}
where $f_{t(b)}$ is the Fermi-Dirac function of top (bottom)
electrode.~\cite{Zwolak2005-1} Room temperature is assumed
throughout this paper and we have assumed that the voltage drops
uniformly in the space between the DNA molecule and electrodes.
Unless otherwise stated, water and ions are not directly included in
the Hamiltonian for transport. Below we do, however, discuss the effect of water and ions
on the HOMO-LUMO gap. We will consider transverse biases
small enough that we do not expect chemical reactions to occur, and
definitely smaller than the electrolysis threshold for water of
about 1.2V.

The justification for using a tight-binding approach is two-fold.
First, the current through the junction is strongly dependent upon
the coupling of the nucleotide (in the junction region) to the
electrodes. Thus, the important quantity to reproduce is the
molecular orbitals of the bases and backbone relative to the
electrodes in order to get the DNA-electrode coupling. We find that
the tight-binding calculations reproduce very well the molecular
orbitals of charge-neutral nucleotides when compared to density
functional theory (DFT) calculations. We have performed DFT
calculations with DZVP basis set and GGA functional B88-PW91. The
HOMO, LUMO, and nearby electronic levels have similar molecular
wavefunctions in both the DFT and tight-binding calculations.
Second, the presence of a nearby counterion (see below) in addition
to the nearby water effectively brings the HOMO-LUMO gap of the
nucleotide close to the one of the charge neutral case (although the
positioning of the gold Fermi level may change with respect to these
states), while the molecular orbitals of the bases remain nearly
identical regardless of the presence of the counterion. However, the
position of the counterion is not fixed, but it fluctuates around
the charge on the backbone at an average distance of 6.5
$\textrm{\AA}$. This will cause additional fluctuations in the
current across the junction. Nonetheless, since the current value is
mainly controlled by the DNA-electrode coupling, these fluctuations
are less likely to contribute than those due to structural motion.

To analyze the dynamics of the DNA strand as it propagates through
the pore, classical molecular dynamics simulations have been
performed with the package NAMD2.~\cite{Kale1999-1} For the
interaction in between the DNA, water, and ions the CHARMM27 force
field~\cite{Foloppe2000-1,MacKerell2000-1} was used, while UFF
parameters~\cite{Rappe1992-1} were used for interactions between the
Si$_3$N$_4$ membrane and other atoms. The location of the
Si$_3$N$_4$ atoms was assumed to be frozen throughout the
simulation\footnote{This also prevents the system from drifting due
to the external electric field.}. When integrating over time, a time
step of 1 fs was used and the temperature of the system was kept
constant at 300K throughout the whole simulation by coupling all but
the hydrogen atoms to a thermal bath with a Langevin damping
constant of 0.2 $ps^{-1}$. The van der Waals interactions were
gradually cut off starting 10 $\textrm{\AA}$ from the atom until
reaching zero interaction 12 $\textrm{\AA}$ away. The DNA strand was
driven through the pore with a large electric field of 6 kcal/(mol
$\textrm{\AA}$  e) in order to achieve feasible simulation times.
This field is larger than the ones used experimentally and results
in a negligible stabilizing electric field, $\mathbf{E_\perp}$.
Thus, for sampling current distributions of the nucleotides, we turn
off $\mathbf{E_\parallel}$ when a base is aligned in between the
electrodes. This gives an adequate representation of the structural
fluctuations when $|\mathbf{E_\perp}|\gg
|\mathbf{E_\parallel}|$\footnote{One of our main conclusions is that
the device has to be operated in this regime for sequencing to be
possible.}.

The pore analyzed in this article is made up of a 24 $\textrm{\AA}$
thick silicon nitride membrane in the
$\beta-$phase.~\cite{Grun1979-1} The experimental etching of the
pore is mimicked by removing the atoms inside a double conical shape
with a minimum diameter of 1.4 nm located at the center of the
membrane and with an outer diameter of 2.5 nm (see Figure 1 for a
schematic representation of the pore). This corresponds to a cone
angle of 20 degrees for each of the cones. In actual experimental
realizations of nanopores such geometry may not always be realized.
For the conclusions of this paper this would not matter as the
calibration of the device would correct for geometrical
imperfections in the pore, as discussed later in the paper. A
sphere, with a radius of 6 nm, of TIP3 water~\cite{Jorgensen1981-1}
is placed around the pore and 1M of potassium and chlorine ions are
added. Spherical boundary conditions are used under NVT conditions
and the size of the membrane is chosen slightly smaller than the
water sphere, such that water molecules can just pass through on the
sides of the membrane. Finally, a single-stranded poly(X) (where X
is A, T, C or G) molecule is generated by removing one strand from a
helical double stranded polynucleotide. At the initial time of the
simulation, this molecule is placed parallel to the pore such that
the tip of the single strand is just inside the pore.

\section{RESULTS AND DISCUSSIONS}

We take a two step computational approach to examine the issues
described above: {\it 1)} molecular dynamics simulations are used to
sample real time atomistic coordinates of the DNA, water, and ions
in a prototypical Si$_3$N$_4$ nanopore, and {\it 2)} these
coordinates are used in quantum mechanical calculations to find the
current across the nanostructure (see also the section Setup and
Methods).

We discuss our results in three subsections: In {\it Setup}, we
examine a larger pore diameter and weaker transverse field than in
ref.~\onlinecite{Lagerqvist2006-1}; In {\it Nucleotide
Stabilization}, we look at the current distribution with varying
transverse field strength; In {\it Influence of the Environment}, we
look at the role of the environment on the electronic conductance.

\begin{figure}
%\begin{center}
\includegraphics*[width=8.5cm]{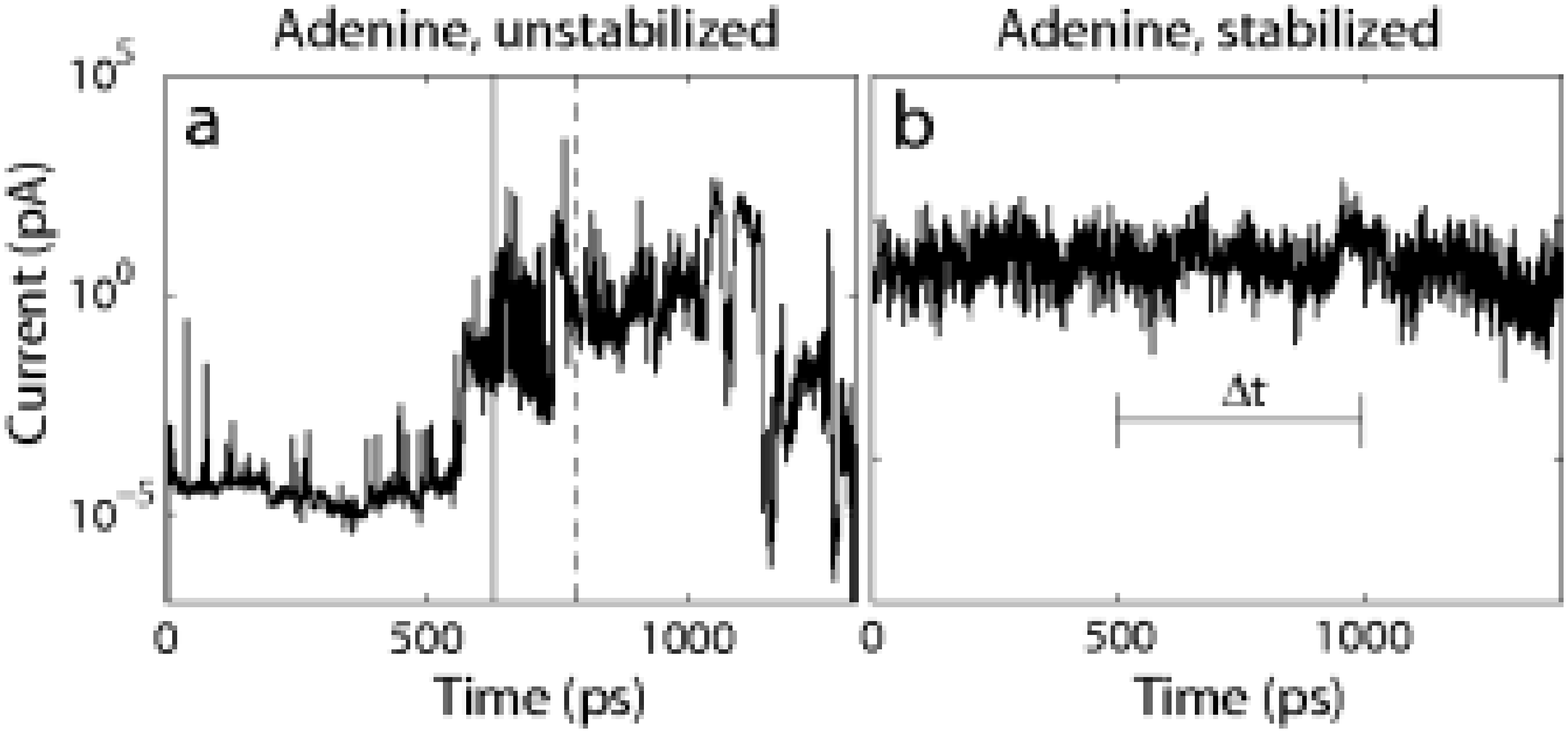}
\caption{} \label{currA}
%\end{center}
\end{figure}

{\it Setup} - Figure \ref{currA}a shows the current as a function of
time as a ss-DNA strand made up of 15 Adenine bases propagates
through a 14 $\textrm{\AA}$  diameter pore\footnote{The distance is
measured from the atomic coordinates of the outermost atoms.} with
embedded electrodes. A pore with a diameter smaller than ~10
$\textrm{\AA}$ does not allow the translocation of ss-DNA, while a
pore with a diameter larger than about 15 $\textrm{\AA}$ causes a
much smaller conductance. Figure \ref{currA}b shows the current as a
function of time when $\mathbf{E_\parallel}$ has been turned off.
The starting condition was taken from the simulation used for Figure
\ref{currA}a at the time indicated by the dashed line in Figure
\ref{currA}a, when a base was aligned in between the electrodes. We
can see that as time progresses the nucleotide stays aligned in
between the electrodes due to the interaction of the nucleotide with
the transverse electric field. This is discussed in further detail
later on in the article. Similar curves have been found for all
other DNA bases. The tunneling current of bare electrodes can be
estimated as $I=\frac{2e^2}{h}T\,V$, where $V$ is the bias,
$T=\exp(-2d\sqrt{2mE/\hbar^2})$, where $e$ is the electron charge,
$d$ the electrode spacing, $m$ the electron mass and $E$ the work
function of gold. For $d=14\ \textrm{\AA}$ and $E=5\ \textrm{eV}$,
we find that the current with vacuum in between the electrodes is
$\sim 0.1$ aA at a bias of 0.1 V, i.e., orders of magnitude lower
than the currents obtained with DNA in between the electrodes.

\begin{figure}
%\begin{center}
\includegraphics*[width=7.5cm]{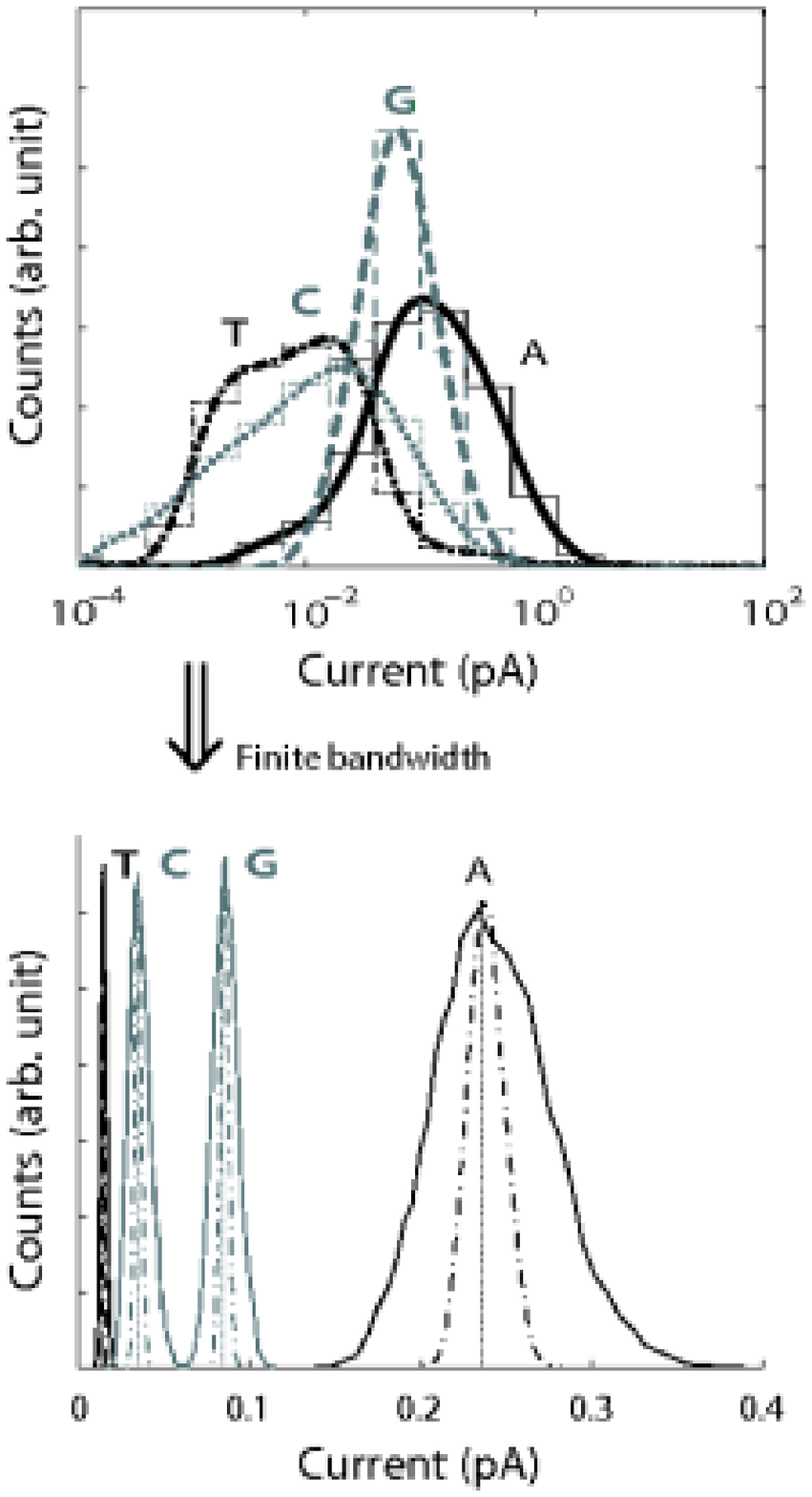}
\caption{} \label{distrATCG}
%\end{center}
\end{figure}

Since we envision operating the nanopore/electrode device in a
regime where the transverse field is much stronger than the driving
field\footnote{Note that it could be possible to operate a device in
this regime as a thermal ratchet, e.g., the thermal fluctuations
overcome the barrier to moving each nucleotide away from the
junction region.}, we can examine the real time structural
fluctuations by sampling the current with the driving field off. The
distributions of these currents, with this particular pore geometry,
for all four bases are shown in the top section of Figure
~\ref{distrATCG}, assuming each current is measured {\it
instantaneously}.\footnote{From the molecular dynamics simulations
we observe that ions fluctuate inside the pore at time scales of the
order of 5 ps due to thermal fluctuations. As the total current
sampling time per base is 1.2 ns, we effectively sample over
multiple ionic configurations.} We can see that these distributions
are unique, but overlapping. This means that a handful of
measurements of a base to be sequenced would not be enough to
distinguish it from the other bases. However, in a real experiment
each measurement would take a finite amount of time to be performed
(finite inverse bandwidth indicated with $\Delta t$ in Figure
~\ref{currA}b). In other words, the electric probes have a finite
bandwidth. Hence, a real measurement averages over a time interval,
which is determined by the sampling frequency of the
electrodes/probes, causing the distributions to narrow around their
average current.

In order to accurately determine the new shape of the distributions
when each measurement is time averaged, one would ideally need
multiple ensembles, where for example one would be the current shown
in Figure ~\ref{currA}b. This is however difficult to realize
numerically. On the other hand, one can assume, as a reasonable
starting point, that the interpolated distributions shown in the top
half of Figure~\ref{distrATCG} are exact. That this is indeed a good
approximation is justified by the following (see also below).
Starting a nucleotide out with the base parallel to the electrode
surfaces, we find that it takes about 100 ps for the transverse
field $|\mathbf{E_\perp}|$ to align it perpendicular to the
electrode surfaces. If one thus waits for a time longer than 100 ps
to sample the distributions, the latter ones must be weakly
dependent on initial conditions. We can then assume that the average
current for another ensemble can be generated by sampling these
interpolated distributions. By repeating this process multiple
times, one can then calculate the new distributions, where each
measurement is {\it time averaged}. These new distributions are
shown in the bottom part of Figure ~\ref{distrATCG}. Clearly, the
new distributions become sharper the more times one samples the
original distributions. The number of samples that should be taken
is determined by the ratio between the period of physical
measurement and the time interval for two measurements to be
considered independent of each other. While the sampling frequency
can be determined exactly, it is hard to give an exact value of the
time needed in between two instantaneous measurements for them to be
considered independent. However, we can take the timescale for
atomic movements in the simulation, which is about 1 ps, as a rough
estimate.

Assuming no external noise, the distributions in the lower half of
Figure ~\ref{distrATCG} show that as long as one samples the
instantaneous distributions at least 100 times (solid lines), the
new distributions will be completely separated from each other, and
if sampled 1000 times the new distributions assume much sharper
shapes. Assuming that the time scale for independent sample
measurements is of the order of 1 ps, one would need an apparatus
sampling at a rate no faster than $\sim 1/(1000\times1
\textrm{ps})=1 \textrm{GHz}$ to obtain disjoint distributions. Since
sampling frequencies of the order of \textrm{GHz} or less are
relatively easy to obtain, we conclude that a {\it single} current
measurement may be sufficient to distinguish the different bases. We
would like to stress again that, in this analysis, we have assumed
no other external source of noise is present (like, for instance,
telegraph noise or $1/f$ noise). If thermal noise is of concern, one
may reduce the sampling frequency, which both reduces the thermal
noise and sharpens the intrinsic current distributions. For example,
at a sampling frequency of 100 kHz, the rms current thermal noise
for Guanine is of the order of 40 fA\footnote{The root mean square
current thermal noise is given by $i_n=\sqrt{4k_bT\Delta f/R}$,
where $T$ is the temperature, $\Delta f$ is the sampling frequency,
and $R$ is the resistance of the nucleotide in the junction.}, while
the distributions due to structural fluctuations assume
$\delta-$function like shapes.\footnote{A larger average current can
be obtained by either increasing the bias or by slightly reducing
the pore size.} If the distributions of the four bases were to begin
overlapping due to external noise, multiple measurements per base
would be needed, and a statistical analysis similar to the one
presented in ref.~\onlinecite{Lagerqvist2006-1} could be performed.
In the rest of the article we will keep on discussing the case where
each measurement is assumed to be instantaneous, as this case
contains more information; one can always transform any distribution
given below into a ``finite-bandwidth'' one as described above.

Current distributions for a different pore diameter and transverse
field were examined previously in
ref.~\onlinecite{Lagerqvist2006-1}, assuming instantaneous
measurements. Although the electrode spacing is larger in this
paper, 14 $\textrm{\AA}$ compared to 12.5 $\textrm{\AA}$, and the
electrode bias is smaller, 0.1 V compared to 1.0 V, the
distributions show remarkable similarities. Adenine shows the
largest mode in both cases and Guanine the second largest. Notice,
however, that in the previous paper Thymine had a slightly larger
mode current than Cytosine, while the results presented here are
reversed. This can be attributed to the change in electrode spacing
and highlights the importance of calibrating the device (see
discussion below). It also demonstrates the importance of the
DNA-electrode coupling: in the upright configuration for which the
nucleotides are held, the Thymine nucleotide has the largest
base-electrode distance, and therefore its coupling to the second
electrode is most sensitive to the electrode spacing.\footnote{For
larger electrode spacing, the nucleotide conductance is strongly
dependent on the base-electrode distance. As the electrode spacing
gets smaller, the current depends more on the spatial character and
energies of the molecular orbitals.} On the other hand the currents
are orders of magnitude larger for the smaller pore/larger bias
case. For example, the mode conductance for Adenine is roughly 1,000
times larger. This difference can be attributed to the change in the
electrode spacing; a smaller pore radius will result in a larger
current.

Following the discussion in the previous paragraph, we would like to
emphasize that it is highly unlikely that the exact geometry
of two pores would be identical. Since in nanoscale systems a single atom change
in the contact geometry or local environment may lead to a substantial change in the
current~\cite{Diventra2000-1,Yang2003-1}, the first step to sequence a DNA
strand would be to calibrate the device by creating these
distributions for the pore at hand. Then a strand can be sequenced
by comparing its current, at each base location, to these target
distributions.

One final comment before we go on to discuss stabilization. When
$\mathbf{E_\parallel}$ is turned off, we observe a mild drift of the
Thymine(T) nucleotide, which causes a slightly different form for
its current distribution when assuming that each measurement is
instantaneous. In our previous work we found nearly Gaussian
distributions for A and G nucleotides but not C and T. Since C and T
are smaller bases, we believe their homogeneous strands have more
freedom to move within the pore. When the strand moves enough, a
nearest neighbor base can come in close vicinity of the electrodes.
For the same reasons, the nearest neighbor base makes a larger
difference in the conductance of C and T bases.\footnote{In a
stretched configuration, the nearest neighbor bases do not affect
the conductance very much so long as the electrode width is on the
order of a single nucleotide.~\cite{Zwolak2005-1} However, the
molecular dynamics simulation allows many configurations to be
explored, including ones where the nucleotide is near the edge of
the electrode and thus the neighboring nucleotide can make a
substantial contribution to the conductance of the junction.} Thus,
a potential future direction of research may be to examine the role
of base sequence on DNA motion within the nanopore.

{\it Nucleotide Stabilization} - Each phosphate group carries a
negative charge in solution. In addition, as we have anticipated, we
find that there is always a counterion fluctuating about the
backbone charge, at an average distance of 6.5 $\textrm{\AA}$. This
helps neutralizing the DNA charge inside the pore as confirmed by
other molecular dynamics simulations~\cite{Rabin2005-1} and also
experiment.~\cite{Sauer-Budge2003-1,Mathe2004-1} However, our
results show that the transverse field can still act as a very
effective stabilizer on the resulting nucleotide-counterion dipole.

\begin{figure}
%\begin{center}
\includegraphics*[width=7.5cm]{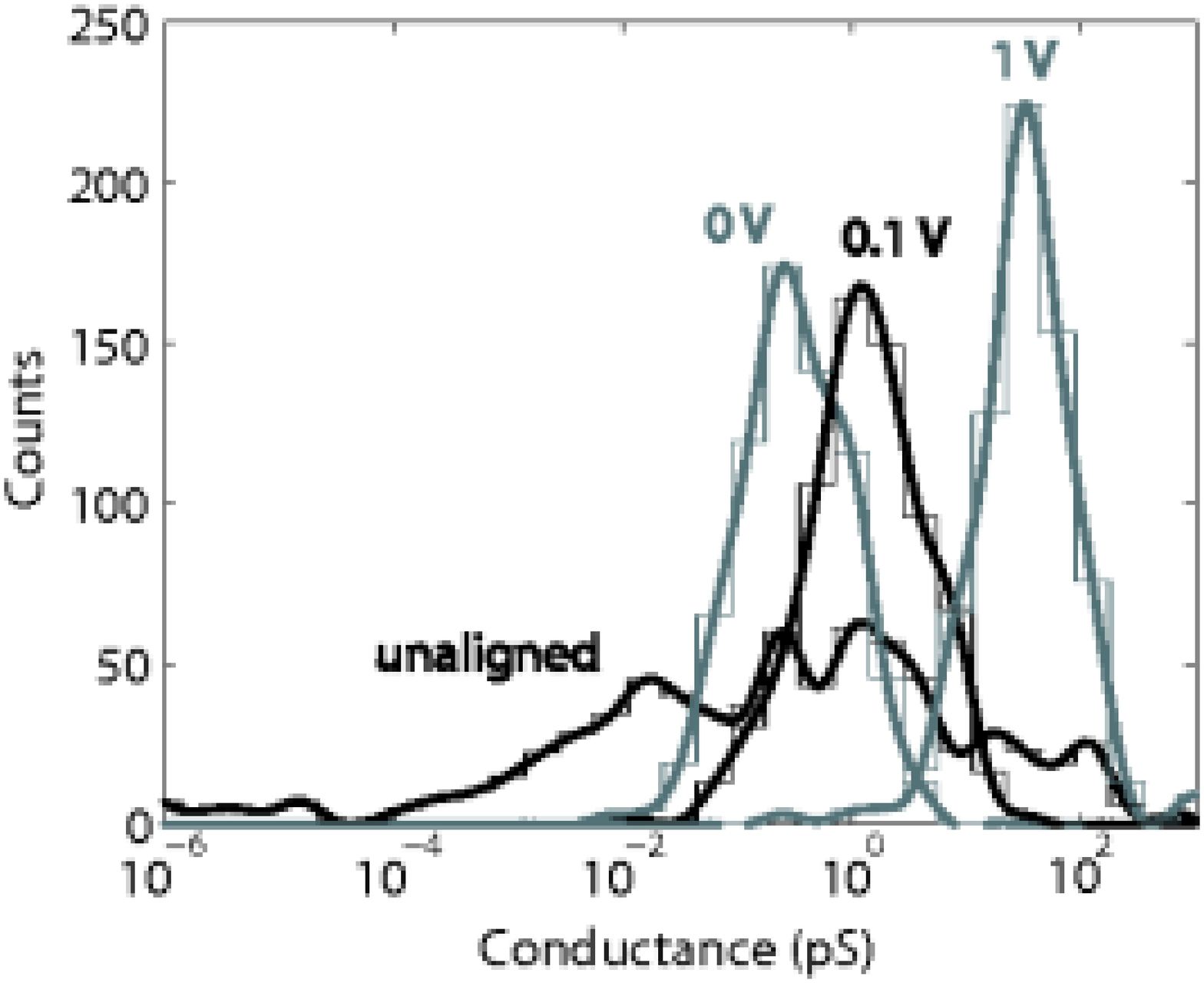}
\caption{} \label{distrVarBias}
%\end{center}
\end{figure}

To better understand the effect of the stabilizing field on the
current, one can compare the conductance distributions for varying
transverse electric fields. In Figure \ref{distrVarBias} the
conductance distributions for four different cases are shown. The
black curve shows the conductance distribution for the completely
unstabilized and unaligned case, generated by transforming the
current in Figure \ref{currA}a into a distribution starting from the
time when the strand starts to translocate. The blue curve is the
conductance distribution generated by turning $\mathbf{E_\parallel}$
off at a time when a base is aligned in between the electrodes, as
indicated by the dashed line in Figure \ref{currA}a, while not
including any stabilizing field. This corresponds to the limiting
case in which both $\mathbf{E_\parallel}$ and $\mathbf{E_\perp}$ are
allowed to approach zero with $|\mathbf{E_\perp}|\gg
|\mathbf{E_\parallel}|$. Red and green distributions are generated
in the same manner as the blue curve, but with both a bias and
stabilizing field of 0.1 V and 1.0 V, respectively. As expected, the
distribution for the unaligned case is much broader, compared to the
other cases, as the bases are oriented in all possible directions
while the strand propagates through the pore with a driving field
much larger than the stabilizing field. When starting with a base
aligned in between the two electrodes the distributions are clearly
much sharper. One also notices an increase of conductance of almost
an order of magnitude for a stabilizing field of 0.1 V compared to
the one with no stabilizing field. Furthermore, the conductance
increases by almost two orders of magnitude when increasing the
stabilizing field from 0.1 V to 1.0 V. This confirms the effect of
the stabilizing field: as the field increases in strength it pulls
the backbone closer to one electrode and aligns the base toward the
other electrode, which increases the conductance. The alignment of
the base with the field is also favored by the steric effect of the
alignment of the backbone with one of the electrodes.

In order to examine in more detail the effect of the stabilizing
field, we have analyzed a case in which the driving field is turned
off as an Adenine base was aligned in the direction {\em
perpendicular} to the stabilizing field. Then, as a bias of 0.1 V is
turned on in between the electrodes, the base starts to align with
the stabilizing field and is completely aligned after about 100
picoseconds. We can thus conclude that as long as the strand is
pulled through the pore at a pace slower than $\sim$100 ps per base,
the stabilizing electric field induced by this bias will be
sufficient for sequencing purposes. In
experiments~\cite{Fologea2005-1,Fologea2005-2} the typical
translocation speed is much slower than the reorientation time of
$\sim$100 ps, allowing the bases sufficient time to reorient with
the transverse electric field.

{\it Influence of the Environment} - As we discussed above, water
and counter-ions do influence the electronic structure of
nucleotides in solution. Without these species, the nucleotides
would have an unscreened charge and electronic transport would be
quite different.

Nevertheless, the fluctuation of water molecules around the
nucleotides inside the pore has little direct effect on the
electrical current. Figure \ref{water} shows the current calculated both with and without
water, for a Guanine base stabilized in between the
electrodes. The presence of water lowers the current on average by
18\% but the additional fluctuations in the current due to transport across water molecules are
negligible compared to the larger structural fluctuations of the DNA molecule.
For the smaller pore geometry used in the previous
paper~\cite{Lagerqvist2006-1}, an even smaller change of 4\% was
observed for Adenine as it propagates through the pore unstabilized.
This is an expected result, as the smaller pore allows for fewer
water molecules to enter, with a corresponding decrease of its
contribution to the current. Also, the small effect of water on the
current is the result of a slight modification of the nucleotide
electronic states. This is opposed to an increase in conductance due
to ``bridging'' effects, which could occur if the pore were
larger.~\cite{Lin2005-1} Considering that structural fluctuations
account for orders of magnitude change in conductance, we can
conclude that the direct effect of water on conductance can be
neglected.

\section{CONCLUSIONS}
By combining molecular dynamics simulations with quantum mechanical
current calculations, we examined the feasibility of DNA sequencing
via transverse electronic transport. Specifically, we have shown
that unless the current is sampled with very high frequencies, i.e.,
larger than $\sim$1 GHz, the time averaging occurring in the probe
apparatus will reduce the fluctuations in the current to such a
level that an individual current measurement may be sufficient to
sequence the base, assuming that any external sources of noise in
the current are small. We also addressed how the transverse field
strength, pore diameter, and local environment affect the
distinguishability of the bases. Furthermore, we have shown that the
electric field induced by the electrodes will effectively stabilize
the bases and hence allows for accurate sequencing when
$|\mathbf{E_\perp}|\gg |\mathbf{E_\parallel}|$. In addition, we have
discussed how ions and water effectively screen the charged
nucleotides. We have also shown that water in the environment has a
negligible direct effect on the electrical current. Our results also
emphasize the importance of device calibration.

It is important to note that there are many issues left to be
addressed, like DNA-surface bonding, time for the DNA to find and go
through the pore (which will depend on device parameters), and the
direct effect of ionic fluctuations on the electronic transport properties
of the DNA.

We thank D. Branton, M. Ramsey and P. Wolynes for useful discussions
and critical reading of our manuscript. This research is supported
by the NIH-National Human Genome Research Institute (JL and MD) and
by the National Science Foundation through its Graduate Fellowship
program (MZ).

\clearpage
\section*{Figure Legends}
\subsubsection*{Figure~\ref{schematic}.}
Schematic of a nanopore (dark gray) with embedded electrodes (light
gray) attached to the edges of the pore. The electrodes are used to
inject a current through the nucleotides in the direction transverse
to the backbone. The electronic signature can then be used to
sequence the DNA.

\subsubsection*{Figure~\ref{currA}.}
Current, at a bias of 0.1 V, as a function of time for
$\mathrm{poly(A)}_\mathrm{15}$, as (a) it is propagating through a
pore with two electrodes without a stabilizing field, and (b) when
the driving field is turned off at a time (indicated by the dashed
line in (a)) while a base is aligned in between the electrode pair.
Solid vertical line in (a) indicates the time at which the DNA
starts propagating through the pore. The transverse electric field
is included in the simulations for figure (b). $\Delta t$ represents
a finite inverse bandwidth.

\subsubsection*{Figure~\ref{distrATCG}.}
Top graph shows the probability distributions, assuming
instantaneous measurements, of currents at a bias of 0.1 V for
$\mathrm{poly(X)}_\mathrm{15}$, where X is A/T/C/G for the solid
black/dashed-dotted black/dotted gray/dotted gray curves,
respectively. The thin lines show the actual current intervals used
for the count, while the thick lines are an interpolation. After the
driving electric field is turned off, the system is left to
equilibrate for 200 ps before samples are taken. Each distribution
is made up of 1200 samples, each taken with a 1 ps interval. The
bottom graph again shows the probability distributions, but now with
the assumption that each measurement is time averaged in between
each sample. The solid/dashed-dotted/dotted line assumes that the
distributions in the top graphs are sampled 100/1,000/$10^7$ times
for each new data point.

\subsubsection*{Figure~\ref{distrVarBias}.}

Probability distributions of the conductance for varying stabilizing
fields for $\mathrm{poly(A)}_\mathrm{15}$. Leftmost curve labeled
"unaligned" corresponds to the completely unstabilized case as shown
in Figure \ref{currA}a, while other curves correspond to various
stabilizing fields when the driving field is turned off. The symbol
0 V corresponds to the case in which the base is aligned in between
the electrode, but there is no stabilizing field.

\begin{figure}
%\begin{center}
\includegraphics*[width=7.5cm]{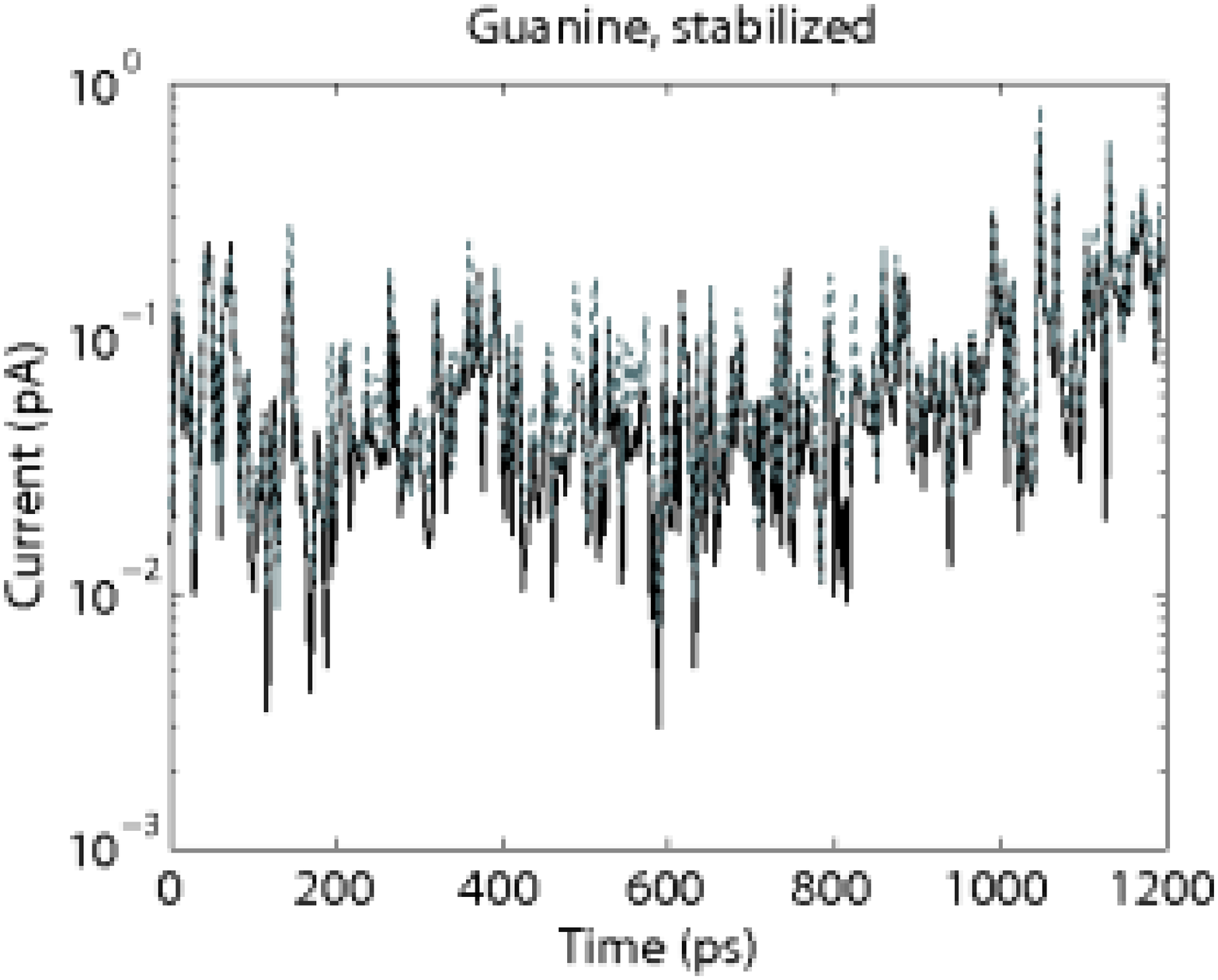}
\caption{} \label{water}
%\end{center}
\end{figure}

\subsubsection*{Figure~\ref{water}.}
Current, at a bias of 0.1 V, as a function of time for
$\mathrm{poly(G)}_\mathrm{15}$, for a single Guanine base stabilized
in between the electrodes by the transverse electric field, with
(solid black curve) and without (dashed gray curve) water included
in the current calculation.

\clearpage

\pagebreak[4]

\pagebreak[4]

\pagebreak[4]

\pagebreak[4]

\section{REFERENCES}

\begin{thebibliography}{42}
\providecommand{\natexlab}[1]{#1}

\bibitem{Lander2001-1}
Lander, E.~S., et~al. 2001.
\newblock Initial sequencing and analysis of the human genome.
\newblock \emph{Nature (London, U. K.)}. 409:860--921.

\bibitem{Venter2001-1}
Venter, J.~C., et~al. 2001.
\newblock The sequence of the human genome.
\newblock \emph{Science (Washington, DC, U. S.)}. 291:1304--1351.

\bibitem{Collins2003-1}
Collins, F.~S., E.~D. Green, A.~E. Guttmacher, and M.~S. Guyer. 2003.
\newblock A vision for the future of genomics research.
\newblock \emph{Nature}. 422:835--847.

\bibitem{Kasianowicz1996-1}
Kasianowicz, J.~J., E.~Brandin, D.~Branton, and D.~W. Deamer. 1996.
\newblock Characterization of individual polynucleotide molecules using a
  membrane channel.
\newblock \emph{Proc. Natl. Acad. Sci. U. S. A.} 93:13770--13773.

\bibitem{Braslavsky2003-1}
Braslavsky, I., B.~Hebert, E.~Kartalov, and S.~R. Quake. 2003.
\newblock Sequence information can be obtained from single DNA molecules.
\newblock \emph{Proc. Natl. Acad. Sci. U. S. A.} 100:3960--3964.

\bibitem{Zwolak2005-1}
Zwolak, M., and M.~Di~Ventra. 2005.
\newblock Electronic signature of DNA nucleotides via transverse transport.
\newblock \emph{Nano Lett.} 5:421--424.

\bibitem{Gracheva2006-1}
Gracheva, M.~E., A.~Xiong, A.~Aksimentiev, K.~Schulten, G.~Timp, and J.-P.
  Leburton. 2006.
\newblock Simulation of the electric response of DNA translocation through a
  semiconductor nanopore--capacitor.
\newblock \emph{Nanotechnology}. 17:622--633.

\bibitem{Lagerqvist2006-1}
Lagerqvist, J., M.~Zwolak, and M.~Di~Ventra. 2006.
\newblock Fast DNA sequencing via transverse electronic transport.
\newblock \emph{Nano Lett.} 6:779--782.

\bibitem{Akeson1999-1}
Akeson, M., D.~Branton, J.~J. Kasianowicz, E.~Brandin, and D.~W. Deamer. 1999.
\newblock Microsecond time-scale discrimination among polycytidylic acid,
  polyadenylic acid, and polyuridylic acid as homopolymers or as segments
  within single RNA molecules.
\newblock \emph{Biophys. J.} 77:3227--3233.

\bibitem{Deamer2000-1}
Deamer, D.~W., and M.~Akeson. 2000.
\newblock Nanopores and nucleic acids: prospects for ultrarapid sequencing.
\newblock \emph{Trends Biotechnol.} 18:147--151.

\bibitem{Meller2000-1}
Meller, A., L.~Nivon, E.~Brandin, J.~Golovchenko, and D.~Branton. 2000.
\newblock Rapid nanopore discrimination between single polynucleotide
  molecules.
\newblock \emph{Proc. Natl. Acad. Sci. U. S. A.} 97:1079--1084.

\bibitem{Vercoutere2001-1}
Vercoutere, W., S.~Winters-Hilt, H.~Olsen, D.~Deamer, D.~Haussler, and
  M.~Akeson. 2001.
\newblock Rapid discrimination among individual DNA hairpin molecules at
  single-nucleotide resolution using an ion channel.
\newblock \emph{Nat. Biotechnol.} 19:248--252.

\bibitem{Meller2001-1}
Meller, A., L.~Nivon, and D.~Branton. 2001.
\newblock Voltage-driven DNA translocations through a nanopore.
\newblock \emph{Phys. Rev. Lett.} 86:3435--3438.

\bibitem{Deamer2002-1}
Deamer, D.~W., and D.~Branton. 2002.
\newblock Characterization of nucleic acids by nanopore analysis.
\newblock \emph{Acc. Chem. Res.} 35:817--825.

\bibitem{Meller2002-1}
Meller, A., and D.~Branton. 2002.
\newblock Single molecule measurements of DNA transport through a nanopore.
\newblock \emph{Electrophoresis}. 23:2583--2591.

\bibitem{Li2003-1}
Li, J., M.~Gershow, D.~Stein, E.~Brandin, and J.~A. Golovchenko. 2003.
\newblock DNA molecules and configurations in a solid-state nanopore
  microscope.
\newblock \emph{Nat. Mater.} 2:611--615.

\bibitem{Nakane2003-1}
Nakane, J.~J., M.~Akeson, and A.~Marziali. 2003.
\newblock Nanopore sensors for nucleic acid analysis.
\newblock \emph{J. Phys.: Condens. Matter}. 15:R1365--R1393.

\bibitem{Aksimentiev2004-1}
Aksimentiev, A., J.~B. Heng, G.~Timp, and K.~Schulten. 2004.
\newblock Microscopic kinetics of DNA translocation through synthetic
  nanopores.
\newblock \emph{Biophys. J.} 87:2086--2097.

\bibitem{Chen2004-1}
Chen, P., J.~Gu, E.~Brandin, Y.-R. Kim, Q.~Wang, and D.~Branton. 2004.
\newblock Probing single DNA molecule transport using fabricated nanopores.
\newblock \emph{Nano Lett.} 4:2293--2298.

\bibitem{Fologea2005-1}
Fologea, D., M.~Gershow, B.~Ledden, D.~S. McNabb, J.~A. Golovchenko, and J.~Li.
  2005.
\newblock Detecting single stranded DNA with a solid state nanopore.
\newblock \emph{Nano Lett.} 5:1905--1909.

\bibitem{Heng2006-1}
Heng, J.~B., A.~Aksimentiev, C.~Ho, P.~Marks, Y.~V. Grinkova, S.~Sligar,
  K.~Schulten, and G.~Timp. 2006.
\newblock The electromechanics of DNA in a synthetic nanopore.
\newblock \emph{Biophys. J.} 90:1098--1106.

\bibitem{Li2001-1}
Li, J., D.~Stein, C.~McMullan, D.~Branton, M.~J. Aziz, and J.~A. Golovchenko.
  2001.
\newblock Ion-beam sculpting at nanometre length scales.
\newblock \emph{Nature (London, U. K.)}. 412:166--169.

\bibitem{Storm2003-1}
Storm, A.~J., J.~H. Chen, X.~S. Ling, H.~Zandbergen, and C.~Dekker. 2003.
\newblock Fabrication of solid-state nanopores with single-nanometre precision.
\newblock \emph{Nat. Mater.} 2:537--540.

\bibitem{Harrell2003-1}
Harrell, C.~C., S.~B. Lee, and C.~R. Martin. 2003.
\newblock Synthetic single-nanopore and nanotube membranes.
\newblock \emph{Anal. Chem.} 75:6861--6867.

\bibitem{Li2004-1}
Li, N.~C., S.~F. Yu, C.~C. Harrell, and C.~R. Martin. 2004.
\newblock Conical nanopore membranes. preparation and transport properties.
\newblock \emph{Anal. Chem.} 76:2025--2030.

\bibitem{Lemay2005-1}
Lemay, S.~G., D.~M. van~den Broek, A.~J. Storm, D.~Krapf, R.~M.~M. Smeets,
  H.~A. Heering, and C.~Dekker. 2005.
\newblock Lithographically fabricated nanopore-based electrodes for
  electrochemistry.
\newblock \emph{Anal. Chem.} 77:1911--1915.

\bibitem{Mannion2006-1}
Mannion, J.~T., C.~H. Reccius, J.~D. Cross, and H.~G. Craighead. 2006.
\newblock Conformational analysis of single DNA molecules undergoing
  entropically induced motion in nanochannels.
\newblock \emph{Biophys. J.} 90:4538--4545.

\bibitem{Biance2006-1}
Biance, A.~L., J.~Gierak, E.~Bourhis, A.~Madouri, X.~Lafosse, G.~Patriarche,
  G.~Oukhaled, C.~Ulysse, J.~C. Galas, Y.~Chen, and L.~Auvray. 2006.
\newblock Focused ion beam sculpted membranes for nanoscience tooling.
\newblock \emph{Microelectron. Eng.} 83:1474--1477.

\bibitem{Ji2006-1}
Ji, Q., Y.~Chen, L.~L. Ji, X.~M. Jiang, and K.~N. Leung. 2006.
\newblock Ion beam imprinting system for nanofabrication.
\newblock \emph{Microelectron. Eng.} 83:796--799.

\bibitem{Kale1999-1}
Kale, L., R.~Skeel, M.~Bhandarkar, R.~Brunner, A.~Gursoy, N.~Krawetz,
  J.~Phillips, A.~Shinozaki, K.~Varadarajan, and K.~Schulten. 1999.
\newblock Namd2: Greater scalability for parallel molecular dynamics.
\newblock \emph{J. Comput. Phys.} 151:283--312.

\bibitem{Foloppe2000-1}
Foloppe, N., and A.~D. MacKerell. 2000.
\newblock All-atom empirical force field for nucleic acids: I. parameter
  optimization based on small molecule and condensed phase macromolecular
  target data.
\newblock \emph{J. Comput. Chem.} 21:86--104.

\bibitem{MacKerell2000-1}
MacKerell, A.~D., and N.~K. Banavali. 2000.
\newblock All-atom empirical force field for nucleic acids: Ii. application to
  molecular dynamics simulations of DNA and RNA in solution.
\newblock \emph{J. Comput. Chem.} 21:105--120.

\bibitem{Rappe1992-1}
Rappe, A.~K., C.~J. Casewit, K.~S. Colwell, W.~A. Goddard, and W.~M. Skiff.
  1992.
\newblock Uff, a full periodic-table force-field for molecular mechanics and
  molecular-dynamics simulations.
\newblock \emph{J. Am. Chem. Soc.} 114:10024--10035.

\bibitem{Grun1979-1}
Grun, R. 1979.
\newblock Crystal-structure of $\beta$-Si$_3$N$_4$ - structural and stability
  considerations between $\alpha$-Si$_3$N$_4$ and $\beta$-Si$_3$N$_4$.
\newblock \emph{Acta Crystallographica Section B-Structural Science}.
  35:800--804.

\bibitem{Jorgensen1981-1}
Jorgensen, W.~L. 1981.
\newblock Pressure-dependence of the structure and properties of liquid n-butane.
\newblock \emph{J. Am. Chem. Soc.} 103:4721--4726.

\bibitem{Diventra2000-1}
Di~Ventra, M., S.~T. Pantelides, and N.~D. Lang. 2000.
\newblock First-principles calculation of transport properties of a molecular
  device.
\newblock \emph{Phys. Rev. Lett.} 84:979.

\bibitem{Yang2003-1}
Yang, Z.~Q., N.~D. Lang, and M.~Di~Ventra. 2003.
\newblock Effects of geometry and doping on the operation of molecular
  transistors.
\newblock \emph{Appl. Phys. Lett.} 82:1938.

\bibitem{Rabin2005-1}
Rabin, Y., and M.~Tanaka. 2005.
\newblock DNA in nanopores: Counterion condensation and coion depletion.
\newblock \emph{Phys. Rev. Lett.} 94:148103.

\bibitem{Sauer-Budge2003-1}
Sauer-Budge, A.~F., J.~A. Nyamwanda, D.~K. Lubensky, and D.~Branton. 2003.
\newblock Unzipping kinetics of double-stranded DNA in a nanopore.
\newblock \emph{Phys. Rev. Lett.} 90:238101.

\bibitem{Mathe2004-1}
Math\'e, J., H.~Visram, V.~Viasnoff, Y.~Rabin, and A.~Meller. 2004.
\newblock Nanopore unzipping of individual DNA hairpin molecules.
\newblock \emph{Biophys. J.} 87:3205--3212.

\bibitem{Fologea2005-2}
Fologea, D., J.~Uplinger, B.~Thomas, D.~S. McNabb, and J.~Li. 2005.
\newblock Slowing DNA translocation in a solid-state nanopore.
\newblock \emph{Nano Lett.} 5:1734--1737.

\bibitem{Lin2005-1}
Lin, J.~P., I.~A. Balabin, and D.~N. Beratan. 2005.
\newblock The nature of aqueous tunneling pathways between electron-transfer
  proteins.
\newblock \emph{Science (Washington, DC, U. S.)}. 310:1311--1313.

\end{thebibliography}
\end{document}